\documentclass[epjCONF]{svjour}

\usepackage{graphics}
\usepackage[varg]{txfonts}
\usepackage[latin1]{inputenc}
\session-title{Ageing low mass stars: from red giants to white dwarfs}
\begin{document}
\title{White dwarf cooling sequences and cosmochronology}
\author{J. Isern\inst{1,2}\fnmsep\thanks{\email{isern@ieec.cat}} \and 
            A. Artigas\inst{3} \and 
            E. Garc\'\i a--Berro\inst{2,4} 
            }
\institute{Institut de Ci\`encies de l'Espai (CSIC), 
           Facultat de Ci\`encies, 
           Campus UAB, 
           Torre C5-parell, 
           08193 Bellaterra, 
           Spain \and 
           Institute for Space Studies of Catalonia,
           c/Gran Capit\`a 2--4, Edif. Nexus 104,   
           08034  Barcelona, Spain \and 
           CRECIM, Universitat Aut\`onoma de Barcelona,
           Campus UAB, 08193 Bellaterra, Spain \and 
           Departament de F\'\i sica Aplicada, 
           Universitat Polit\`ecnica de Catalunya,  
           c/Esteve Terrades, 5,  
           08860 Castelldefels,  
           Spain}
\abstract{The  evolution  of white  dwarfs  is  a simple  gravothermal
  process. This means that  their luminosity function, i.e. the number
  of white dwarfs  per unit bolometric magnitude and  unit volume as a
  function  of  bolometric magnitude,  is  a monotonically  increasing
  function that decreases abruptly as  a consequence of the finite age
  of the  Galaxy. The  precision and the  accuracy of the  white dwarf
  luminosity functions obtained with the recent large surveys together
  with the improved quality of  the theoretical models of evolution of
  white dwarfs allow to feed the hope that in a near future it will be
  possible  to  reconstruct  the  history of  the  different  Galactic
  populations.}
\maketitle
\section{Introduction}
White  dwarfs  represent  the  last  evolutionary  stage  of  low  and
intermediate mass stars, i.e. stars  with masses smaller than $10\pm 2
\, M_\odot$. Most of them are composed of carbon and oxygen, but white
dwarfs with masses smaller than $0.4\, M_\odot$ are made of helium and
are members  of close  binary systems, while  those more  massive than
$\sim 1.05\, M_\odot$ are probably  made of oxygen and neon. The exact
composition  of the  cores  of carbon-oxygen  white dwarfs  critically
depends on the evolution during  the previous red giant and asymptotic
giant branch  phase, and more specifically on  the competition between
the carbon--$\alpha$ and triple-$\alpha$  reactions, on the details of
the  stellar evolutionary  codes and  on the  choice of  several other
nuclear cross  sections.  In a typical  case --- for  instance a white
dwarf of  $0.58\, M_\odot$ ---  the total amount of  oxygen represents
the  62\% of the  total mass  while its  concentration in  the central
layers of the white dwarf can be as high as 85\%.

In all  cases, the core is surrounded  by a thin layer  of pure helium
with a mass ranging from  $10^{-2}$ to $10^{-4}\, M_\odot$. This layer
is, in  turn, surrounded by an  even thinner layer of  hydrogen with a
typical mass lying in the  range of $10^{-4}$ to $10^{-15}\, M_\odot$.
This  layer  is  missing  in  $\sim  25\%$  of  the  cases.  From  the
phenomenological point  of view, white dwarfs  containing hydrogen are
classified as DA while the remaining ones (the collectively denoted as
non-DA) are  classified as DO, DB,  DQ, DZ and DC,  depending on their
spectral features.  The origin of  these spectral differences  and the
relationship among  them has not  been yet elucidated, although  it is
related to  the initial conditions during the  AGB evolutionary phase,
and also to delicate interplay between several physical process, among
which  we  mention  gravitational  and  thermal  diffusion,  radiative
levitation,  convection at  the  H-He and  He-core interfaces,  proton
burning, stellar winds and mass accretion from the interstellar medium
--- see, for instance Fontaine (2013), this volume.

The  structure  of  white  dwarfs  is sustained  by  the  pressure  of
degenerate  electrons  and  these  stars  cannot  obtain  energy  from
thermonuclear reactions.  Therefore, their evolution  can be described
in  terms of  a  simple  cooling process  \cite{mest52}  in which  the
internal degenerate core  acts as a reservoir of  energy and the outer
non-degenerate   layers  control   the  energy   outflow.    A  simple
calculation indicates  that the time  they take to fade  and disappear
beyond the capabilities of the  present telescopes is very long, $\sim
10$  Gyr.  Thus  the  populations  of white  dwarfs  retain  important
information about the past history  of our Galaxy. In particular, they
allow to obtain  the age of the different  Galactic components, namely
the disk, the  spheroid and the system of  open and globular clusters,
as  well  as   the  star  formation  history  of   the  Galactic  disk
\cite{alth10,font01,hans03,iser98,koes02,koes90}.

The tool to obtain such  information is the luminosity function, which
is defined  as the number  of white dwarfs  of a given  luminosity per
unit volume and magnitude interval:

\begin{equation}
N(l) = \int^{M_{\rm s}}_{M_{\rm i}}\,\Phi(M)\,\Psi[T-t_{\rm cool}(l,M)-t_{\rm PS}(M)]
\tau_{\rm cool}(l,M) \;dM
\label{ewdlf}
\end{equation}
\noindent

\noindent where  $T$ is the  age of the  population under study,  $l =
-\log  (L/L_\odot)$,  $M$  is  the   mass  of  the  parent  star  (for
convenience all  white dwarfs  are labeled with  the mass of  the main
sequence  progenitor), $t_{\rm  cool}$  is the  cooling  time down  to
luminosity   $l$,    $\tau_{\rm   cool}=dt/dM_{\rm   bol}$    is   the
characteristic cooling time, $M_{\rm s}$ is the maximum mass of a main
sequence star  able to produce a  white dwarf, and $M_{\rm  i}$ is the
minimum mass of the main sequence  stars able to produce a white dwarf
of  luminosity $l$,  i.e.  is  the mass  that satisfies  the condition
$T=t_{\rm cool}(l,M) + t_{\rm PS}(M)$ and $t_{\rm PS}$ is the lifetime
of the progenitor  of the white dwarf.  The  remaining quantities, the
initial  mass  function,  $\Phi(M)$,  and  the  star  formation  rate,
$\Psi(t)$,  are not  known a  priori  and depend  on the  astronomical
properties  of the  stellar population  under study.  Since  the total
density  of white dwarfs  of a  given population  is not  usually well
known,  to  compare   the  theoretical  and  observational  luminosity
functions  it  is  customary  to  normalize  the  computed  luminosity
function to the bin with the smallest error bar. For instance, for the
case of the  disk white dwarf luminosity function  it is traditionally
chosen  to  normalize at  $l=3$  \cite{iser98}.   In  summary, if  the
observed luminosity  function and  the evolutionary behavior  of white
dwarfs are well  known, it is possible to obtain the  age and the star
formation rate of the evolution under study.

\section{The observed luminosity functions}

\begin{figure}
\vspace*{9.2 cm}
\begin{center}
\includegraphics{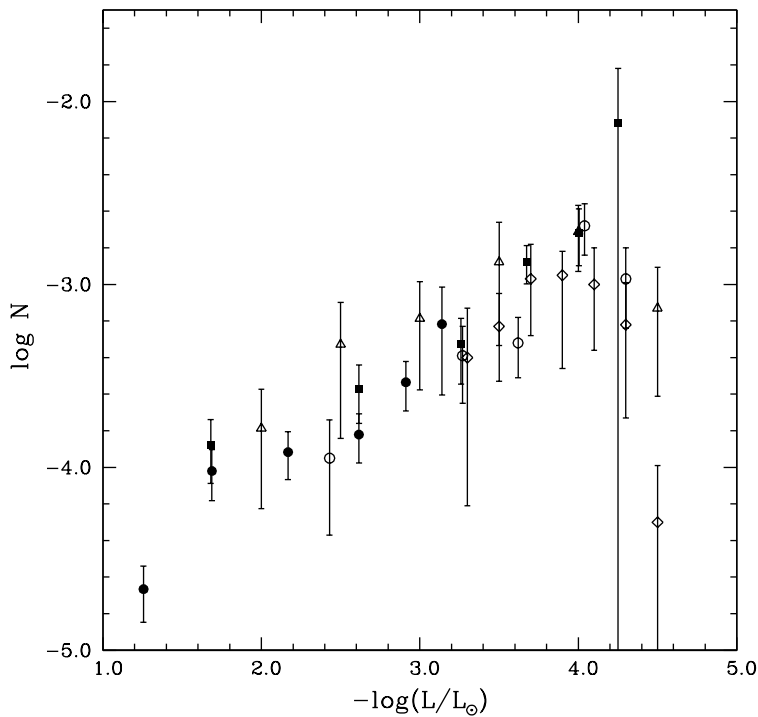}
\caption{Luminosity  functions  obtained   before  the  era  of  large
  surveys.  The different  symbols represent different determinations:
  full  circles   \cite{lieb88},  full  squares   \cite{evan92},  open
  triangles \cite{ossw96},  open diamonds \cite{legg98},  open circles
  \cite{knox99}.}
\end{center}
\label{oldlf}
\end{figure}

The  first luminosity  function  was derived  about  four decades  ago
\cite{weid68}, and since then it has been noticeably improved with the
work of many authors --- see Fig.  \ref{oldlf}. The monotonic behavior
of this function clearly proves  that the evolution of white dwarfs is
a  simple  gravothermal  process,  while  the  sharp  cut-off  at  low
luminosities is the  consequence of the finite age  of the Galaxy. The
recent availability of  data from the Sloan Digital  Sky Survey (SDSS)
has noticeably improved the  accuracy of the new luminosity functions,
as many  new white  dwarfs were  added to the  very limited  sample of
white   dwarfs  with  measured   magnitudes,  parallaxes   and  proper
motions.  For  instance,  the   white  dwarf  luminosity  function  of
Ref.~\cite{harr06} (HA-LF) was built from  a sample of $\sim 6,000$ DA
and non-DA  white dwarfs with  accurate photometry and  proper motions
obtained  from the  SDSS  Data  Release 3  and  the USNO-B  catalogue,
whereas  that of  Ref.~\cite{dege08}  (DG-LF) was  constructed from  a
sample  of   $3,528$  spectroscopically-identified  DA   white  dwarfs
obtained from the SDSS Data Release 4.

The discrepancies between the HA-LF  and the DG-LF at low luminosities
are well  understood, and  can be attributed  to the different  way in
which the effective temperatures and  gravities of the sample of white
dwarfs were observationally determined \cite{dege08}. Furthermore, the
DG-LF only considers  DA white dwarfs and, at  low temperatures, it is
difficult to separate them from  non-DA white dwarfs.  For this reason
in  this work  we will  restrict ourselves  to analyze  the  DG-LF for
magnitudes smaller than $M_{\rm bol} \sim 13$.

\begin{figure}
\vspace*{8 cm}
\begin{center}
\includegraphics{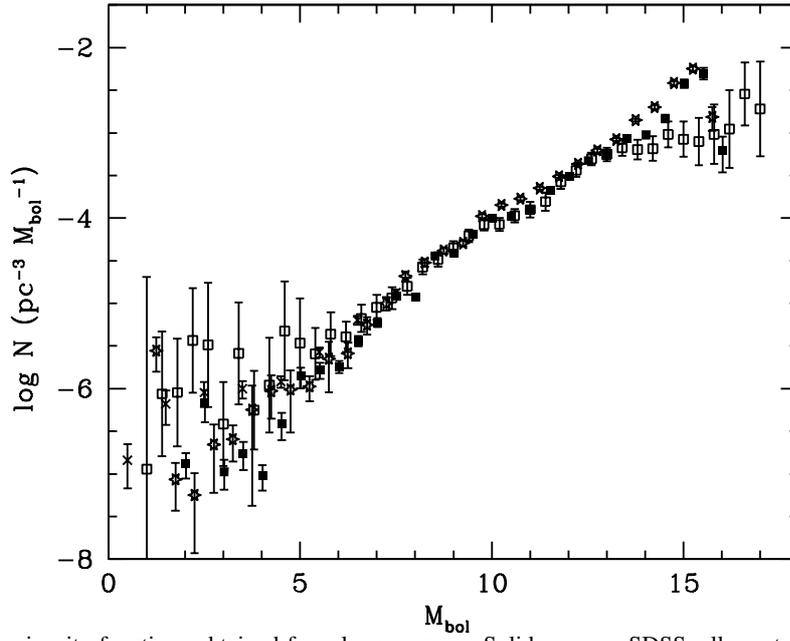}
\caption{Luminosity  functions  obtained  from large  surveys.   Solid
  squares:  SDSS, all  spectral types  \cite{harr06},  hollow squares:
  SDSS,  only DA  white dwarfs  \cite{dege08}, crosses:  SDSS,  hot DA
  white   dwarfs   \cite{krze09},   stars:  SuperCOSMOS   Sky   Survey
  \cite{rowe11}.}
\end{center}
\label{newlf}
\end{figure}

At high  luminosities, say magnitudes  smaller than $M_{\rm  bol} \sim
6$, the dispersion of both  luminosity functions is very large --- see
Fig. \ref{newlf}.   The reason is that both  luminosity functions have
been  built  using  the  reduced  proper motion  method  that  is  not
appropriate  for bright  white  dwarfs.  The  UV-excess technique  has
allowed  to build  a luminosity  function in  the range  $-0.75$  to 7
(KZ-LF) \cite{krze09}.  This method,  however, is not adequate for dim
stars and becomes rapidly incomplete  out of this range of magnitudes.
Fortunately,  this  sample  overlaps  with  the  HA-LF  and,  assuming
continuity, it  is possible to  extend the luminosity function  to the
brightest region.

One  of the potential  problems of  the luminosity  functions obtained
from the SDSS results from the fact that the integration time is fixed
and,  consequently, the  S/N ratio  depends on  the brightness  of the
source. This can  lead to large uncertainties in  the determination of
the  parameters of faint  white dwarfs,  and may  introduce systematic
errors \cite{limo10}. Fortunately, a completely independent luminosity
function  has been  obtained  \cite{rowe11} from  the SuperCOSMOS  Sky
Survey, which culls white dwarfs  using their proper motion. As can be
seen in  Fig.~\ref{newlf}, this luminosity  function (hereafter called
RH-LF)   does  not  excessively   deviate  from   the  HA-LF   at  low
luminosities, thus providing some hope that these luminosity functions
are  not affected by  large systematic  effects.  However,  for bright
white dwarfs this luminosity  function suffers from the same drawbacks
as  the HA-LF  and the  DG-LF luminosity  functions.  Thus,  at bright
luminosities is better to use the KZ-LF data.

Because of  the overlap in the velocity  distributions, the luminosity
function obtained  from reduced  proper motion methods  are in  fact a
superposition of  thin and  thick disc objects.  It has  been recently
shown  \cite{rowe11} that, in  principle, it  is possible  to separate
both populations  using kinematic arguments.   This technique yielded,
for the  first time,  luminosity functions for  both populations  in a
self-consistent  way,   thus  offering  the   possibility  to  provide
interesting  insight  in  the  sequence  of events  that  led  to  the
formation of the galactic disk --- see below.

\section{An overall view of white dwarf cooling}

Since white  dwarfs are degenerate objects, they  cannot obtain energy
from nuclear burning.  Therefore, their evolution can be considered as
a simple cooling process.  Globally, the evolution of their luminosity
can be written as:
\begin{equation}
L+L_{\nu} =- \int^{M_{\rm WD}}_0 C_{\rm v}\frac{dT}{dt}\,dm
         - \int^{M_{\rm WD}}_0 T\Big(\frac{\partial P}{\partial
             T}\Big)_{V,X_0}\frac{dV}{dt}\,dm
          +\;\; (l_{\rm s}+e_{\rm s}) \dot{M}_{\rm s}           
\end{equation}
\label{eener}
where  the l.h.s.   of the  equation represents  the sinks  of energy,
photons  and neutrinos,  while  the r.h.s.   contains  the sources  of
energy, the  heat capacity  of the star,  the compressional  work, the
contribution of the latent heat  release and of the energy released by
gravitationl  settling   upon  crystallization,  times   the  rate  of
crystallization, $\dot{M}_{\rm s}$ \cite{iser98}. This equation has to
be complemented with a  relationship connecting the temperature of the
core with the luminosity of the star.

The evolution of  white dwarfs from the planetary  nebula phase to its
disappearance can be roughly divided in four stages:

\textbf{Neutrino cooling:} The range  of luminosities of this phase is
$\log (L/L_\odot) > -1.5$.  This  stage is very complicated because of
the dependence  on the  initial conditions of  the newly born  star as
well as  on the complex  and not yet  well understood behavior  of the
hydrogen envelope.  If the hydrogen  layer is smaller than  a critical
value,  $M_{\rm H}  \le 10^{-4}$  $M_\odot$, nuclear  burning  via the
pp--reactions quickly drops  as the star cools down  and never becomes
dominant.  Since  astero-seismological observations seem  to constrain
the size  of $M_{\rm H}$ well  below this critical  value, this source
can  be  neglected.    Fortunately,  when  neutrino  emission  becomes
dominant, the  different thermal structures converge to  a unique one,
granting the uniformity of the models with $\log (L/L_\odot)\le -1.5$.
Furthermore,  since the  time necessary  to reach  this value  is $\le
8\times 10^7$ years for any  model, its influence in the total cooling
time  is   negligible  \cite{dant89},   except  of  course   at  large
luminosities.

\textbf{Fluid cooling:}  This phase  occurs at luminosities  $-1.5 \ge
\log  (L/L_\odot)  \ge  -3$.   The   main  source  of  energy  is  the
gravothermal  one.  Since  the  plasma is  not  very strongly  coupled
($\Gamma  <   179$),  its   properties  are  reasonably   well  known.
Furthermore, the flux of energy  through the envelope is controlled by
a thick non degenerate layer with an opacity dominated by hydrogen (if
present) and  helium, and weakly  dependent on the metal  content. The
main source of uncertainty is related to the chemical structure of the
interior,    which   depends    on   the    adopted   rate    of   the
$^{12}$C$(\alpha,\gamma)^{16}$O reaction and on the treatment given to
semiconvection  and overshooting.  If  this rate  is high,  the oxygen
abundance  is higher  in the  center than  in the  outer  layers, thus
resulting  in a reduced  specific heat  at the  central layers  of the
star, where  the oxygen abundance can  be as high  as $X_{\rm O}=0.85$
\cite{sala97}.

\begin{figure}
\vspace*{9 cm}
\begin{center}
\includegraphics{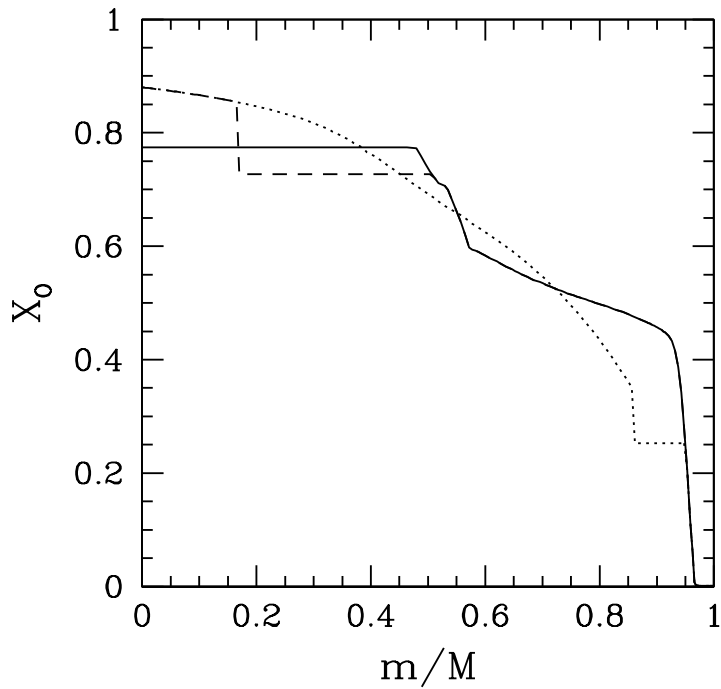}
\caption{Oxygen  profile of a  $0.61\, M_{\odot}$  white dwarf  at the
  beginning of  solidification (solid line)  and when the  $\sim 15\%$
  (dashed  line) and  $\sim 85\%$  (dotted line)  in mass  has already
  crystallized.}
\end{center}
\label{fig3}
\end{figure}

\textbf{Crystallization:}  White dwarfs with  $\log (L/L_\odot)  < -3$
are expected to  experience a first order phase  transition, and their
deep  cores   crystallize  at  these   luminosities.   Crystallization
introduces two  new sources of energy: latent  heat and sedimentation.
In the case of Coulomb plasmas, the latent heat is small, $\sim k_{\rm
B} T_{\rm s}$ per nuclei,  where $k_{\rm B}$ is the Boltzmann constant
and $T_{\rm s}$ is the temperature of solidification. Its contribution
to  the total  luminosity is  small,  $\sim 5$\%,  but not  negligible
\cite{shav76}.   During the  crystallization process,  the equilibrium
chemical compositions of  the solid and liquid plasmas  are not equal.
Therefore, if  the resulting solid  flakes are denser than  the liquid
mixture, they  sink towards the  central region. If they  are lighter,
they rise upwards and melt where the solidification temperature, which
depends  on the  density,  becomes  equal to  that  of the  isothermal
core. The  net effect is a  migration of the  heavier elements towards
the  central  regions with  the  subsequent  release of  gravitational
energy \cite{moch83,iser97}. Of course, the efficiency of the process depends
on  the detailed  chemical  composition and  on  the initial  chemical
profile and it  is maximum for a mixture made of  half oxygen and half
carbon  uniformly distributed  through  all the  star \cite{iser00}.  An  additional
source  of  energy   that  has  to  be  taken   into  account  is  the
gravitational  diffusion  of $^{22}$Ne  synthesized  from the  initial
content of $^{12}$C, $^{14}$N and $^{16}$O during the He-burning phase
\cite{garc08}.

\textbf{Debye cooling:}  When almost all the star  has solidified, the
specific  heat follows Debye's  law. However,  the outer  layers still
have very  large temperatures  as compared with  the Debye's  one, and
since their  total heat capacity  is still large enough,  they prevent
the sudden disappearance  of the white dwarf at least  for the case of
white dwarfs with thick hydrogen envelopes \cite{dant89}.

\section{Cooling sequences}

In this work  we have adopted the BASTI  models \footnote{These models
can    be    downloaded    from    http://www.oa-teramo.inaf.it/BASTI}
\cite{sala10}. These models can  follow the diffusion of the different
chemical species, convective mixing,  residual nuclear burning and all
the  phenomena  related  with  the crystallization  process,  and  the
evolutionary  ages  are  in  excellent  agreement  with  other  recent
calculations \cite{rene10}.  The  parameters for the envelopes adopted
here are  $q_{\rm He} = 10^{-2}  M_{\rm WD}$ and $q_{\rm  H} = 10^{-4}
M_{\rm WD}$  for the DAs and  $q_{\rm He} = 10^{-3.5}  M_{\rm WD}$ for
the non-DAs.

The choice of the chemical  composition of the white dwarf interior is
of critical  importance since all the factors  influencing the cooling
rate, specific  heat, neutrino emission,  crystallization temperature,
sedimentation and so on, depend on the detailed chemical structure. In
the  present work,  the chemical  profile has  been  obtained assuming
solar  metallicity, convective overshooting  during the  main sequence
and  semiconvection  during central  He-burning,  while the  breathing
pulses  occurring  at  the  end  of  the  core  He-burning  have  been
suppressed    \cite{sala10}.     The     adopted    rate    for    the
$^{16}$C$(\alpha,\gamma)$$^{16}$O was that of Ref.~\cite{kunz02}.

\begin{figure}
\vspace*{8 cm}
\begin{center}
\includegraphics{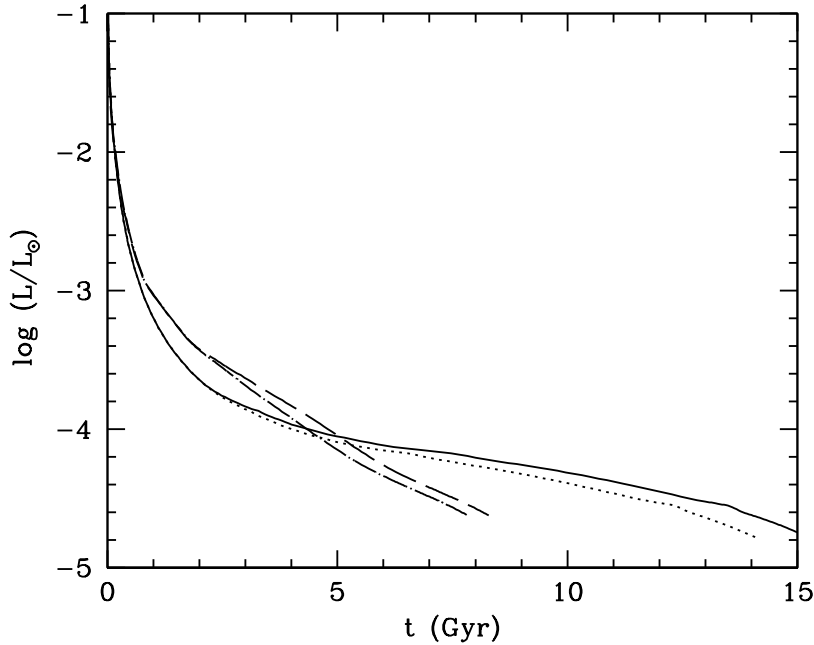}
\caption{Time  evolution of  the  luminosity of  a $0.61\,  M_{\odot}$
  white dwarf. The solid line corresponds to a DA model in which phase
  separation  was included. The  dotted line  corresponds to  the same
  model  but  disregarding  phase   separation.  The  dashed  and  the
  dotted-dashed  lines represent,  respectively, a  non-DA  model when
  separation is included and disregarded.}
\end{center}
\label{fig4}
\end{figure}

Fig.~\ref{fig3}  displays the  oxygen profiles  for the  CO core  of a
$\sim 0.6\, M_\odot$  white dwarf progenitor obtained just  at the end
of the  AGB phase  (solid line). The  inner part  of the core,  with a
constant abundance of $^{16}$O, is determined by the maximum extension
of the  central He-burning convective region while  beyond this region
the oxygen profile is built  when the thick He-burning shell is moving
towards   the  surface.   Simultaneously,   gravitational  contraction
increases its temperature and density and, since the ratio between the
$^{16}$C$(\alpha,\gamma)$$^{16}$O and the  $3\alpha$ reaction rates is
smaller  for higher  temperatures, the  oxygen mass  fraction steadily
decreases  in  the  external   part  of  the  CO  core  \cite{sala97}.
Fig.~\ref{fig3}  also  displays  how  the inner  abundance  of  oxygen
gradually increases in the  inner regions as the crystallization front
advances in mass. The region  with a flat chemical profile placed just
above the  solidification front is  due to the  convective instability
induced upon crystallization. The first calculation of a phase diagram
for CO mixtures was done  several years ago \cite{stev80} and resulted
in an eutectic shape. This  result was a consequence of the assumption
that the  solid was  entirely random. Later  on \cite{segr94},  it was
found that  the CO phase diagram  was of the spindle  form. Because of
this, the solid  formed upon crystallization is richer  in oxygen than
the liquid and  therefore denser. For a $0.6\,  M_{\odot}$ white dwarf
with equal  amounts of  carbon and oxygen,  $\delta \rho  /\rho \simeq
10^{-4}$. Therefore the solid settles down at the core of the star and
the  lighter liquid  left behind  is redistributed  by Rayleigh-Taylor
instabilities  \cite{stev80,moch83,iser97,alth12}.  The  result is  an
enrichment of  oxygen in the central  layers and its  depletion in the
outer ones. Even when rotation  is considered, convection is indeed an
efficient mechanism to redistribute the carbon rich fluid out from the
crystallization front and  that the liquid phase can  be considered as
well mixed \cite{iser97}.

Finally, we mention that  the characteristic cooling time that appears
in Eq.~\ref{ewdlf} not only depends  on the internal energy sources or
sinks but also on the photon luminosity which, in turn, depends on the
transparency  of   the  envelope.    Since  non-DA  models   are  more
transparent than the  DA ones, they cool down much  more rapidly as it
can be seen in Fig.~\ref{fig4}.

\section{The age of the Galactic disk}

\begin{figure}
\vspace*{8 cm}
\begin{center}
\includegraphics{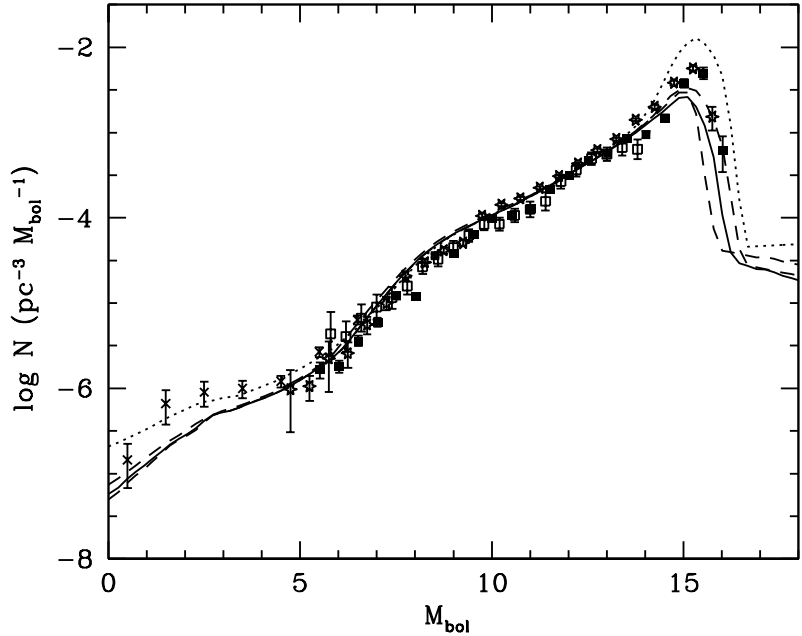}
\caption{White dwarf luminosity  function for different star formation
  rates  and  different  ages  of  the  Galactic  disk.   The  symbols
  corresponding   to    the   observational   data    are   those   of
  Fig.~\ref{newlf}. Dashed  lines, from  left to right,  constant SFR,
  and $t_{\rm disk}  = 10, \, 13$~Gyr. Solid line,  $ \Psi \propto (1+
  \exp[(t-t_0)/\tau])^{-1}$, $\tau=3$~Gyr, $t_0= 10$ and $t_{\rm disk}
  = 13$ Gyr.  Dotted line, $\Psi \propto \exp(-t/\tau)$, $\tau=3$~Gyr,
  $t_{\rm disk} = 13$~Gyr.}
\end{center}
\label{fig5}
\end{figure}

A common  picture of  the formation  of the Milky  Way is  that spiral
galaxies form  as a consequence of  the gas cooling  inside a spinning
dark matter halo.  In a first  stage, the gas collapses in a dynamical
time  scale that  lasts  for several  hundred  million years,  leaving
behind a  spherical stellar  halo, and settles  down into a  disk from
where  stars  form. Galactic  discs  are  structures  that are  easily
destroyed by mergers with other structures of similar mass. Therefore,
if a disk appears almost undamaged, it means that its has not suffered
strong mergers since it was born and that its life has been reasonably
quiet. It  is also possible  for disks formed  early in the life  of a
spiral  galaxy to have  been heated  by minor  mergers leading  to the
formation of a  thick disk able to produce a new  thin disk within it.
According  to this  picture, the  sequence  of events  leading to  the
formation of the  presently observed structure of the  Milky Way could
have  been  the  following  one  \cite{reid05}: i)  Formation  of  the
primitive halo at $t \sim 12  - 13$ Gyr, ii) Episodes of minor mergers
of satellite systems at $t \sim 11-12$ Gyr, iii) Formation of the disk
at $t \sim 10-11$ Gyr, iv)  A major merger produces the formation of a
thick disk at $ t \sim 9-10$ Gyr, and v) Formation of the thin disk at
$ t \sim 8 $ Gyr.

\begin{figure}
\vspace*{8 cm}
\begin{center}
\includegraphics{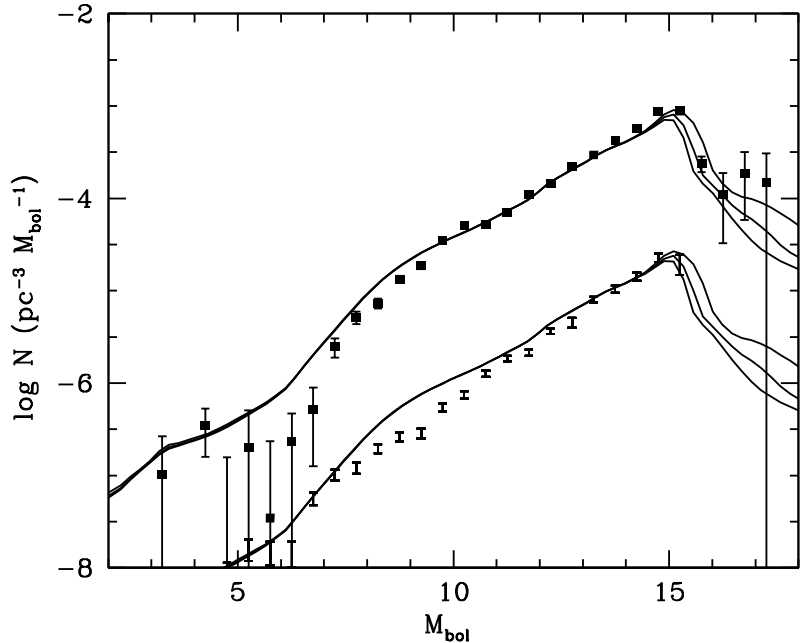}
\caption{White dwarf  luminosity function  of the thin  (upper curves)
  and thick (lower curves) disks \cite{rowe11}. The data corresponding
  to the thick disk have been shifted by a quantity of $-1$ for a sake
  of  clarity.  The theoretical  function  have  been  computed for  a
  constant star  formation rate  and ages of  10, 11 and  13~Gyr (from
  left to right respectively).}
\end{center}
\label{fig6}
\end{figure}

In the case of the halo,  the age is essentially determined by the age
of the system  of globular clusters, which at  present is estimated to
be  $\sim 13$~Gyr. In  the case  of the  disk, several  indicators are
used. The  main ones  are the ages  of F  and G stars,  the luminosity
function of white dwarfs,  radioactive clocks are also often employed,
and finally  the ages  of old  open clusters. Some  of these  ages are
obtained from objects in the solar  neighborhood, as it is the case of
white dwarfs,  assuming they are  representative of the  total Galaxy,
which  is  not the  case.   On the  contrary,  old  open clusters  are
distributed   all   over  the   Galaxy,   and   are  considered   more
representative but, since their  lifetime is relatively short (several
$10^8$ years) many of them could have been destroyed and, in fact they
only provide  a lower limit to the  age of the disk,  for which reason
the local indicators continue to be extremely useful. Incidentally, it
is worth noting that one of the key points in the determination of the
age of the disk is NGC~6791, a very old, extremely metal-rich Galactic
cluster.  The age  of this  cluster estimated  from  the main-sequence
turn-off  method  was  different  from  the  one  estimated  from  the
termination  of the  white dwarf  cooling sequence.   When  the energy
release  due  to the  gravitational  diffusion  of  $^{22}$Ne and  the
settling  of oxygen  upon crystallization are included,
both ages coincide and turn to be 8~Gyr \cite{garc10}.

There  are two important  properties of  the luminosity  function that
deserve a  comment. The  first one is  that, after  normalization, the
bright  part of the  luminosity function  $( M_{\rm  bol} \le  14)$ is
almost  independent of  the star  formation rate  \cite{iser09}, $N(l)
\propto  \langle\tau_{\rm  cool}\rangle$,   unless  a  burst  of  star
formation occurred very recently. The second important fact that needs
to be considered is  that Eq.~(\ref{ewdlf}) do not satisfies Piccard's
theorem  for  the inversion  of  integral  equations.  Thus, the  star
formation  rate cannot  be directly  obtained, as  the unicity  of the
solution  cannot be  guaranteed and  the final  result depends  on the
trial  function  used to  fit  observations.   The  first property  is
clearly illustrated in Fig.~\ref{fig5}, where the luminosity functions
obtained using different SFRs are  almost coincident in the region $ 6
\le  M_{\rm bol}  \le  14$. The  corollary  is that  the BASTI  models
reproduce  reasonably  well the  evolution  of  white  dwarfs in  this
region.

The age of the disk depends on the form adopted for the star formation
rate. If a constant rate is  adopted, the cutoff is compatible with an
age of $\sim 13$ Gyr, i.e. the disk would had been formed a short time
after the primitive halo. If an exponentially declining star formation
rate is  adopted, it  is necessary to  reduce the  age of the  disk to
$\sim 11$ Gyr (not shown in  the figure) to adjust the position of the
cut-off. A  good fit can also  be obtained if an  almost constant rate
lasting  for  the last  $\sim  10$~Gyr  preceded  by an  exponentially
growing star  formation activity is adopted.  This is the  same to say
that the disk started to form from the center to the periphery. Notice
that the cool end of  the luminosity function still shows an important
dispersion in the values and to elucidate among these possibilities it
will be necessary to improve its accuracy at low luminosities.

Obviously, the determination  of the ages of the  thin and thick disks
would be extremely helpful  to prove the previously described sequence
of events.  Fig.~\ref{fig6} displays the observed luminosity functions
of the  thin and thick disks,  as well as  the theoretical predictions
assuming a constant  star formation rate and different  ages. The most
striking feature is  that both structures look as  if they were coeval
since the maximum of both distributions lies approximately at the same
magnitude, $M_{\rm  bol} \sim 15$. Furthermore,  both populations seem
to  be rather  old, $\sim  11$~Gyr, in  agreement with  the luminosity
functions  obtained using  the SDSS.   Certainly, it  is  premature to
extract conclusions  since the cut-off of  the thick disk  is not well
defined and the  modeling of very old white dwarfs  is still plenty of
uncertainties.  Nevertheless, if this  result turns out to be correct,
it  could be  obviously  interpreted  as if  there  was no  difference
between the time at which thin and the thick disk formed, and moreover
that this  unique disk  formed quite  soon in life  of the  Milky Way.
This,  if correct,  could give  support to  the recent  discovery that
sub-populations  with  similar  [$\alpha$/Fe]$-$[Fe/H] have  a  smooth
distribution of scale heights,  thus suggesting that effectively there
is not a  distinctive thick disk population.  It  is also important to
notice  here that  in deriving  the  luminosity function  it has  been
assumed that no  vertical and radial migration of  stars is effective.
If these effects were important,  it would be necessary to compute the
luminosity function in the  context of a complete numerical simulation
of the Galaxy.

\section{Conclusions}

The  use   of  white  dwarfs  as   cosmochronometers  has  experienced
noticeable advances  during the last  years both from  the theoretical
and observational  point of views, and  has become a  reliable tool to
measure the age  of an ensemble population of  stars if the conditions
are  well  defined,  as  is  the  case,  for  instance,  of  NGC~6791.
Furthermore, having for the  first time separated luminosity functions
of the thin  and thick disk opens new  possibilities to understand the
origin and evolution of the Milky Way.

However, several unsolved problems  still remain. From the theoretical
point  of  view, there  are  still  noticeable  differences among  the
different cooling  tracks at  low luminosities. These  differences are
probably due to the use of different boundary conditions \cite{rohr12}
and  different  sizes  and   physics  adopted  for  the  envelope.  An
additional problem  is our incomplete understanding of  the origin and
evolution  of the  DA, non-DA  character, which  could  introduce some
uncertainties  in  the determinations  of  the theoretical  luminosity
function.   From the  observational  point of  view  the main  problem
resides in the still poor  determination of the luminosity function of
cool white dwarfs, as well  as in the criteria to efficiently separate
the  different populations  (thin disk,  thick disk  and  spheroid) of
Galactic white dwarfs.

\vspace{0.1cm}
\noindent  {\sl   Acknowledgements.}   This  research   was  partially
supported by  MICINN grants AYA2011--24704, and  AYA2011--23102 by the
ESF EUROGENESIS  project (MICINN grants EUI2009--04167  and 04170), by
the European Union FEDER funds and by the AGAUR.

\bibliographystyle{epj}

\bibliography{isern}

\end{document}